\documentclass[mathptm]{aipproc}
\layoutstyle{8x11single}

\usepackage{graphicx}
\usepackage{dcolumn}
\usepackage{bm}

\begin{document}

\title{Loop Quantum Cosmology corrections on gravity waves produced during primordial inflation}

\author{J. Grain}{address={Institut d'Astrophysique Spatiale, Universit\'e Paris-Sud 11, CNRS \\ B\^atiments 120-121, 91405 Orsay Cedex, France}}

\begin{abstract}
Loop Quantum Gravity (L.Q.G.) is one of the two most promising tentative theory for a quantum description of gravity. When applied to the entire universe, the so-called Loop Quantum Cosmology (L.Q.C.) framework offers microscopical models of the very early stages of the cosmological history, potentially solving the initial singularity problem via bouncing solutions or setting the universe in the appropriate initial conditions for inflation to start, via a phase of super-inflation. More interestingly, L.Q.C. could leave a footprint on cosmological observables such as the Cosmic Microwave Background (CMB) anisotropies. Focusing on the modified dispersion relation when holonomy and inverse-volume corrections arising from the L.Q.C. framework are considered, it is shown that primordial gravity waves generated during inflation are affected by quantum corrections. Depending on the type of corrections, the primordial tensor power spectrum is either suppressed or boosted at large length scales, and strongly departs from the power-law behavior expected in the standard scenario.
\end{abstract}


\maketitle

\section{Introduction}
Building a complete quantum theory of gravitation is one of the greatest challenges in theoretical physics since the advent of general relativity and quantum mechanics. Since many decades, different proposals emerge and among the theories willing to reconcile Einstein gravity and quantum mechanics, Loop Quantum Gravity (L.Q.G.) is appealing as it is based  on a non-perturbative quantization of the 3-space geometry \cite{rovelli1}. Loop Quantum Cosmology (L.Q.C.) is a finite, symmetry reduced model of L.Q.G.  suitable for the study of the whole Universe as a physical system (see, {\it e.g.}, \cite{bojo0}).

Such a challenge is currently a pure theoretical open question as all the observed phenomena do not {\it a priori} need such a quantum theory of gravity to be explained (at least alternatives to a quantum-gravity-based explaination still exist). However, it is well admitted that even in the absence of observational facts to guide theoretical investigations, very high energetic phenomena can be used not only as {\it gedanken experiments} to test theoretical proposals, but also as future probes of quantum gravitational physics. With black holes evaporation, primordial cosmology is probably the most powerful place in the cosmos where quantum gravity play an important role. 

In its very early stages, our universe went through a phase of cosmic inflation \cite{linde}. Although still debated, this high energetic phenomena (the energy scale of inflation could be as high as $10^{16}$~GeV and most of the inflationary models are inspired by high energy/particle physics or by quantum gravity) has received strong support from many experiments, including from the WMAP 5-Years results \cite{wmap}, and by solving many theoretical issues in cosmology. This exponential growth of the scale factor allows cosmologists of the eighties to solve most of the problems of the non-inflationary Big-Bang models as such an additionnal phase explains {\it e.g.} the curvature of the universe or its homogeneity. Moreover, it provides a mechanism to generate cosmological perturbations in the primordial universe. In most of the inflationary models, only scalar and tensor perturbations are produced. On the one hand, scalar perturbations are the primordial seeds for galaxies and large scale structures formation by gravitational collapse and they leave their footprint on the temperature and E-mode polarization anisotropies of the Cosmic Microwave Background (CMB). On the other hand, tensor perturbations correspond to primordial gravity waves and could be observed via their unique footprint on the undetected yet B-mode anisotropies of the CMB. Different inflationary scenarios, some of them inspired or emerging from quantum gravity settings, {\it a priori} lead to different statistical properties for this cosmological perturbations, which inevitably lead to different statistical properties of the CMB anisotropies and large scale structures. Inversely, those cosmological observables can therefore be used as probes of inflation making primordial cosmology a laboratory to test high energetic and/or quantum gravity proposals.

Current observations of the CMB anisotropies or of the large scale structures are not precise enough to distinguish between different inflationary scenarios. But many new experiments\cite{cmb-exp} such as the PLANCK satellite\cite{planck} or the EBEX balloon-borne experiment\cite{ebex}, are planned, or already flying to make precise measurements of the statistical properties of both temperature and polarization anisotropies of the CMB. In the perspective of this new class of experiments and taking the fact that quantum gravity is now in the playground of cosmology, it becomes promising to study the impact of quantum gravity proposals on the generation of cosmological perturbations during inflation. 

Following this guideline, we consider the influence of L.Q.C. corrections to general relativity on the production of  gravitational waves during inflation. Quite a lot of work has already been devoted to gravitational waves in L.Q.C. \cite{L.Q.C.gen} and in this paper, we follow the path adopted in \cite{grainL.Q.G.2,grainL.Q.G.3,grainL.Q.G.4}. Our approach assumes the background to be described by the standard slow-roll inflationary scenario whereas L.Q.C. corrections are taken into account to compute the propagation of tensor modes within such an inflationary background. This approach is heuristically justified (to decouple the physical effects) and intrinsically plausible (as, on the one hand, the L.Q.C.-driven superinflation can only be used to set the proper initial conditions to a standard inflationary stage if the horizon {\it and} flatness problems are both to be solved \cite{tsuji} and as, on the other hand, it seems that the quantum bounce can trigger on a standard inflationary phase \cite{jakub}).

\section{A non-specialist summary of L.Q.G. and L.Q.C.}
The program of Loop Quantum Gravity is to built a non-perturbative quantization of general relativity and can be viewed as a trial to answer the following question, as formulated in \cite{smolin}:
\begin{quote}
	"Can we construct a quantum theory of spacetime based only the experimentally well confirmed principles of general relativity and quantum mechanics?"
\end{quote}
As expressed in Rovelli's introductory book \cite{rovelli1}, Loop Quantum Gravity considers as seriously as possible the principles of general relativity {\it and} quantum mechanics. In particular, this means that the theory have to be {\it background independant}, or preserve diffeomorphism invariance in a more mathematical spelling.The first step of this program consists in formulating general relativity in a {\it tractable} Hamiltonian framework, which is achieved by making use of the Ashtekar variables as the two canonically conjuguate variables:
\begin{equation}
\begin{array}{ll}
	\displaystyle\mathrm{the~densitized~triad:}&\displaystyle E^a_i\equiv\left|\det(e^b_j)\right|^{-1}e^a_i, \\
	\displaystyle\mathrm{the~Ashtekar~connection:}&\displaystyle A^i_a\equiv\Gamma^i_a+\gamma K^i_a.
\end{array}
\end{equation}
In the above, $e^a_i$ stands for the triad components, $\Gamma^i_a$ for the spin connection and $K^i_a$ for the extrinsic curvature. Finally, $\gamma$ is the so-called Barbero-Immirzi parameter, a real-valued number allowing quantization to be performed on a compact group. (Historically, the Ashtekar variables were making use of a pure imaginary Barbero-Immirzi parameter, $\gamma=i$, allowing the Hamiltonian to be a polynomial functional of the two canonically conjuguate variables. However, the precise value of $\gamma$ does not matter at a classical level as the same equations are recovered for ay values of such a parameter.) In an Hamiltonian framework, the two canonically conjuguate variables are related by a Poisson bracket and the dynamics is thus deduced from the Hamiltonian acting on these canonical variables
\begin{equation}
H_{grav}[N]=\frac{1}{16\pi G} \int_{\Sigma} d^3 x N |\det E|^{-\frac{1}{2}} E_j^a E_k^b (\epsilon_{ijk} F_{ab}^i - 2(1+\gamma^2) K_{[b}^i K_{a]}^j ), \label{Hgr}
\end{equation}
where $F_{ab}^i = \partial_a A_b^i - \partial_b A_a^i + \epsilon^{ijk} A_a^j A_b^k$ is the field strength\footnote{We underline that if the 3-space metric and the extrinsec curvature are used as the two canonical variables instead of the Ashtekar variables, the Hamiltonian constraint leads to the Wheeler-de Witt equation.}. (One easily sees that the above Hamiltonian becomes polynomial in $E^a_i$ and $A^i_a$ by setting $\gamma=i$.)

Such a formulation is based on a $3+1$ decomposition of the metric, {\it i.e.}
\begin{displaymath}
	ds^2=N^2dt^2-q_{ab}\left(dx^a+N^adt\right)\left(dx^b+N^bdt\right)
\end{displaymath}
(with $q_{ab}=e^i_ae_{ib}$ the metric on the 3-hypersurfaces, $N$ the lapse function and $N^a$ the shift vector), which {\it a priori} appears as contradictory with the background independance of the theory. However, this choice is only adopted for convenience as it simplifies the quantization process, and the spacetime foliation can be changed by adopting a different lapse function and shift vector. The background independance therefore requires some constraints to be fulfilled. First, the diffeomorphism constraint ensures the theory to be independant of the shift vector and the space geometry therefore does not depend on the spatial coordinates. Second, the Hamiltonian constraint ensures the theory not to depend on the lapse function and therefore, to be invariant under a different choice of the {\it temporal} coordinates\footnote{Denoting the 3-space foliation as {\it spatial} slices and the coordinates perpendicular to them as the {\it temporal} coordinate should not be understood literally. This notation is convenient but {\it does not} assume a preferred time and general covariance is preserved.}. The general spacetime covariance of the theory is consequently preserve if those two constraints are satisfied. Finally, the theory have to be invariant under any rotations of the triad fields, resulting in a third contraint called the Gauss constraint.

At a quantum level, the Poisson bracket relating the two canonical variables are replaced by a commutation relation. However, in standard quantum field theory, the quantization algebra is well defined only after integrating over the 3-space, therefore assuming a pre-defined background. Such a procedure cannot be adopted if one want to preserve the background independance of the theory ensuring all the geometrical variables to be dynamical ones. As a consequence, the program of L.Q.G. consists secondly in quantizing in terms of two new canonical and covariant variables: the holonomy of the Ashtekar connection and the flux of the densitized triad
\begin{equation}
\begin{array}{ccc}
	\displaystyle h_e(A)=\mathcal{P}\exp\left(\int_e\tau_iA^i_a\dot{e}^ad\lambda\right)&\displaystyle\mathrm{and}&\displaystyle F_S(E)=\int_S\tau_iE^i_an^ad^2y.
\end{array}
\end{equation}
The Poisson algebra of those quantities is well known and their representation on a Hilbert space can therefore be looked for. Following this program, the kinematical Hilbert space of the theory has been found: it possesses an orthonormal basis $\left|\{\Gamma\}\right>$ labelled by the embeddings of the spin-network in the manifold. It can be understood as the set of eigenvectors of geometrical operators such as area or volume operators, both of them possessing a discrete and finite spectra. The complete quantization of the theory is however not accomplished yet as the dynamical Hilbert space still needs to be found by solving the Hamiltonian constraints (the spinfoam approach could be a possible way to solve this problem).

Nevertheless, the theory of Loop Quantum Gravity already enjoys many successes in solving different open issues in theoretical physics. Among them, one can particularly notice that singularities in general relativity (precisely where Einstein theory breaks down) are removed. Because of the {\it granular} structure of spacetime, ultraviolet divergences in quantum field theory are removed. This theory also explains the entropy of black holes in terms of the statistical mechanics of the states associated with the horizon degrees of freedom. Finally, this framework has been applied to the case of cosmology, leading to the developpement of a non-perturbative quantum theory of our universe, {\it a priori} valid up to the Planck scale.

Loop Quantum Cosmology is a FLRW-reduced formulation of Loop Quantum Gravity (see \cite{bojo0,calcagni} and reference therein for an introductory review of L.Q.C.). The background cosmological equations can re-derived in the Ashtekar formalism by restricting to the homogeneous case and using conformal time (in the following, overdot means conformal-time derivative, prime means cosmic-time derivative and overbar are for homogeneous variables)
\begin{equation}
	ds^2=a^2\left(d\eta^2-dx^idx_i\right),
\end{equation}
and
\begin{eqnarray}
	\bar{E}^a_i=\bar{p}\delta^a_i&\mathrm{and}&\bar{A}^a_i=\gamma\bar{k}\delta^a_i, \\
	\bar{N}^a=0&\mathrm{and}&\bar{N}=\sqrt{\bar{p}}
\end{eqnarray}
with $\bar{p}=a^2$ and $\bar{k}=\gamma\dot{a}$, leading to the following gravitational Hamiltonian
\begin{equation}
	\bar{H}^{(cl)}_{grav}\left[N\right]=\frac{1}{16 \pi G}\displaystyle\int_\Sigma d^3x\bar{N}\left[-6\sqrt{\bar{p}}\bar{k}^2\right]=-\frac{3}{8\pi G}\displaystyle\int_\Sigma d^3x \left(\gamma a\dot{a}\right)^2.
\end{equation}
Combined with the matter Hamiltonian, a scalar field $\Phi$ in the following, and using the Hamilton-Jacobi equation for general relativity, we can recover the standard Friedman equation and Klein-Gordon equation in a FLRW background
\begin{equation}
\begin{array}{l}
	\displaystyle\left(\frac{\dot{a}}{a}\right)^2=a^2\frac{8\pi G}{3}\rho_\phi, \\
	\displaystyle\ddot{\Phi}+2\frac{\dot{a}}{a}\dot\Phi+a^2\frac{\partial V(\Phi)}{\partial\Phi}=0.
\end{array}
\end{equation}

In this paper, we will only consider the first order quantum corrections coming from the use of holonomy on the one hand and from the use of densitized triad (also called inverse-volume correction) on the other. This latter term is somehow more speculative than the former one as it was shown to exhibit a fiducial cell dependence (see, {\it e.g.}, \cite{ashtekar4}). For the sake of completeness it is however obviously worth considering the fully corrected propagation of gravitational waves. Those first order corrections can be implemented in a quasi-classical framework as summarized in \cite{calcagni}. The full L.Q.C. theory is first quantized. From the recovered Hilbert space, it is then possible to define some quasi-classical states, the analog of coherent states in quantum optics. They can be viewed as wavepackets concentrated around the classical trajectory. It is then possible to defined a quasi-classical Hamiltonian containing first order quantum corrections by simply computing the expectation value of the Hamiltonian operator over those quasi-classical states
\begin{equation}
	\bar{H}^{(qc)}_{grav}\left[\bar{N}\right]=\frac{1}{16 \pi G}\displaystyle\int_\Sigma d^3xS(\bar{p})\bar{N}\left[-6\sqrt{\bar{p}}\left(\frac{\sin\bar{\mu}\gamma\bar{k}}{\bar{\mu}\gamma}\right)^2\right]
\end{equation}
for the gravitational sector and
\begin{equation}
	H^{(qc)}_{matter}=\displaystyle\int_\Sigma d^3x\left(\frac{1}{2}D(\bar{p})\frac{\dot{\Phi}^2}{\bar{p}^{3/2}}+\bar{p}^{3/2}V(\Phi)\right)
\end{equation}
for the matter sector. The first set of corrections comes from holonomies and are encoded in the $\left(\frac{\sin\bar{\mu}\gamma\bar{k}}{\bar{\mu}\gamma}\right)$ term. The $\bar\mu$ quantity can be interpreted as the operator providing the size of loop used to define the holonomy. The second set of corrections comes from inverse-volume and are encoded in the $S$ and $D$ functions. If the universe is in a semi-classical regime, they can be parametrized as 
\begin{equation}
	\alpha(\bar{p})=1+\lambda\left(\frac{L^2_{Pl}}{\bar{p}}\right)^{\kappa/2},
\end{equation}
with $\lambda$ and $\kappa$ two positive numbers, and where $\alpha$ stands for either $S$ or $D$. From those effective, quasi-classical Hamiltonians, one can derive some effective background equations which contain first order quantum corrections
\begin{equation}
\begin{array}{l}
	\displaystyle\left(\frac{\dot{a}}{a}\right)^2=a^2\frac{8\pi G}{3}\rho_\phi\left(S(a)-\frac{\rho_\Phi}{\rho_c}\right), \\
	\displaystyle\ddot{\Phi}+2\frac{\dot{a}}{a}\left(1-\frac{1}{2}\frac{a}{\dot{a}} \frac{\dot{D}}{D} \right)\dot\Phi+a^2D(a)\frac{\partial V(\Phi)}{\partial\Phi}=0,
\end{array}
\end{equation}
with $\rho_c=3/8\pi G \gamma^2\bar\mu^2a^2$ a critical energy density which cannot be exceeded. This quadratic form of the Friedman equation is at the origin of the quantum bounce which replaces the Big-Bang singularity. In addition to that, because of the $(1-\frac{1}{2}\frac{a}{\dot{a}} \frac{\dot{D}}{D})$ term or because of a negative valued Hubble parameter during a potential contracting phase before the bounce, the Klein-Gordon equation now exhibits an {\it anti-friction} term. As consequence, before or during the quantum phase, this anti-friction term makes the scalar field to climb up its potential. At the end of the quantum phase, the scalar field could then be in the appropriate conditions for a standard inflationary phase to start \cite{tsuji,jakub}. As a consequence, L.Q.C. corrections allow the quantum universe filled with a scalar field to reach the appropriate state for a standard inflationary cosmology to begin.

\section{The loopy tensor power spectrum}
The same quasi-classical approach can be used to investigate the influence of Loop Quantum Cosmology on the generation of cosmological perturbations in the early universe. Following \cite{bojo1}, the first step consists in defining the classical dynamics of the perturbations. If one restricts to the case of gravitational waves, the perturbed metric and Ashtekar variables reads :
\begin{equation}
	ds^2=a^2\left(d\eta^2-\left(\delta_{ij}+h_{ij}\right)dx^idx^j\right),
\end{equation}
and
\begin{eqnarray}
	E^a_I=\bar{E}^a_i+\delta E^a_i=\bar{E}^a_i-\frac{1}{2}\bar{p}h^a_i&\mathrm{and}&K^i_a=\bar{K}^i_a+\delta K^i_a=\bar{K}^i_a+\frac{1}{2}\dot{h}^i_a+\frac{1}{2}\bar{k}h^i_a.
\end{eqnarray}
The classical perturbed Hamiltonian is now given by
$$
H^{(cl)}_{grav}\left[N\right]=\bar{H}^{(cl)}_{grav}\left[\bar{N}\right]+\delta H^{(cl)}_{grav}\left[\delta E,\delta K\right] 
$$
with
\begin{displaymath}
	\delta H^{(cl)}_{grav}=\frac{1}{16\pi G}\int_{\Sigma}\mathrm{d}^3x \bar{N} \left[ - \frac{\bar{k}^2}{2\bar{p}^{3/2}} (\delta E^c_j\delta E^d_k\delta_c^k\delta_d^j) + \sqrt{\bar{p}} (\delta K_c^j\delta K_d^k\delta^c_k\delta^d_j)- \frac{2 \bar{k}}{\sqrt{\bar{p}}} (\delta E^c_j\delta K_c^j) 
- \frac{1}{\bar{p}^{3/2}} (\delta_{cd} \delta^{jk}  E^c_j \delta^{ef} \partial_e \partial_f  E^d_k ) 
\right]. 
\end{displaymath}
As is the case for the background dynamic, the dynamic of perturbations is given by the Hamilton-Jacobi equations 
\begin{eqnarray}
	\delta\dot{E}^a_i&=&\left\{\delta{E}^a_i,H^{(cl)}_{grav}+H^{(cl)}_{matter}\right\}, \\
	\delta\dot{K}^a_i&=&\left\{\delta{K}^a_i,H^{(cl)}_{grav}+H^{(cl)}_{matter}\right\}, 
\end{eqnarray}
from which one can deduce the standard equation of motion for gravitational waves in a FLRW bcakground :
\begin{displaymath}
	\ddot{h}^a_i+2\frac{\dot{a}}{a}\dot{h}^a_i-\nabla^2{h}^a_i=0.
\end{displaymath}
Switching to spatial Fourier space and multiplying the gravity waves by the scale factor, {\it i.e.} $\phi_k(\eta)=a(\eta)\int d^3x {h}^a_i(\eta,\vec{x})e^{i\vec{k}\vec{x}}$, we derived the Schr\"odinger equation commonly used in inflationary cosmology:
\begin{equation}
	\ddot{\phi}_k+\left(k^2-\frac{\ddot{a}}{a}\right)\phi_k=0.
\end{equation}

\subsection{Holonomy corrections}
\begin{figure}
	\includegraphics[scale=0.42]{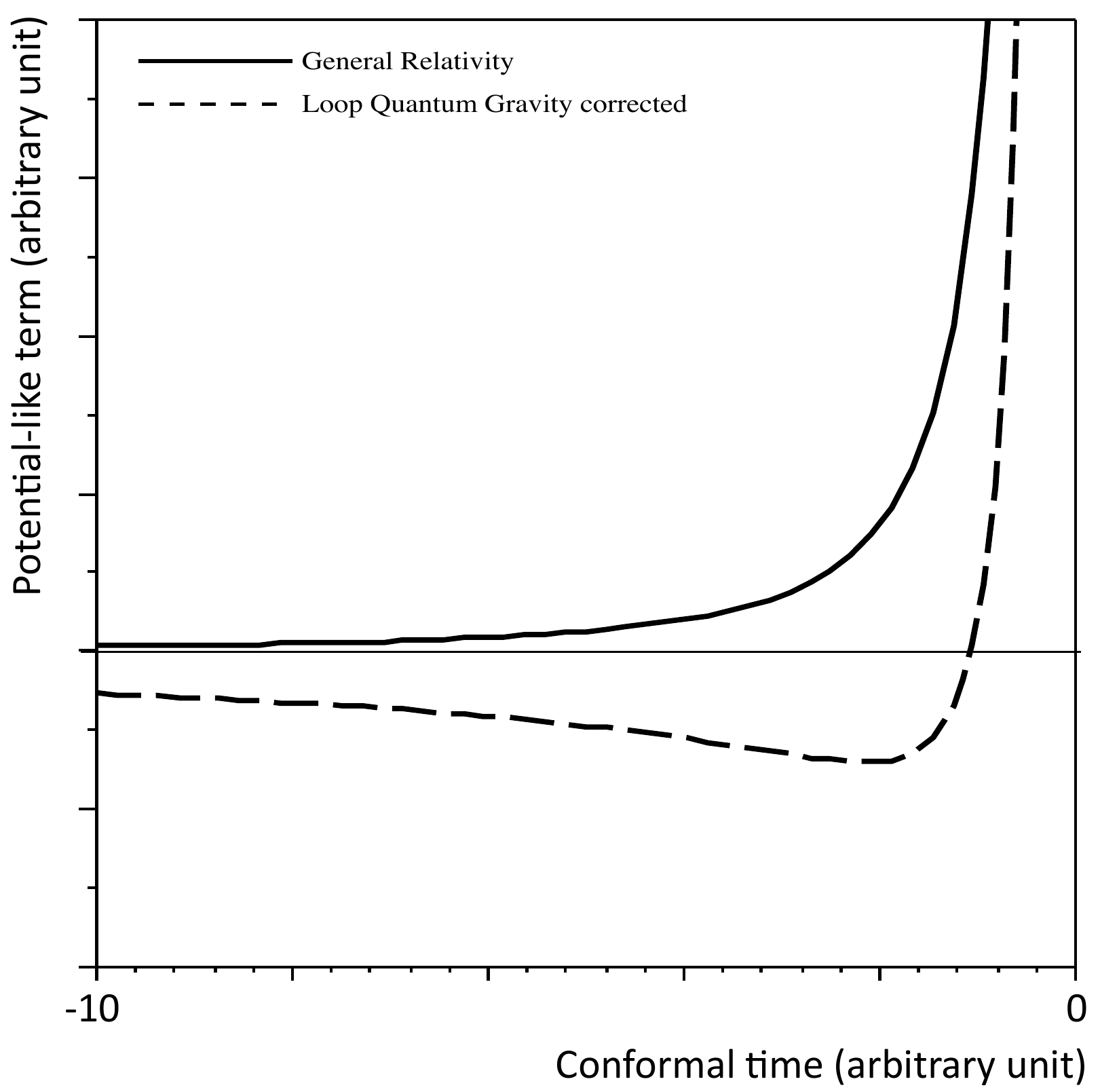}\hspace{1.cm}\includegraphics[scale=0.4]{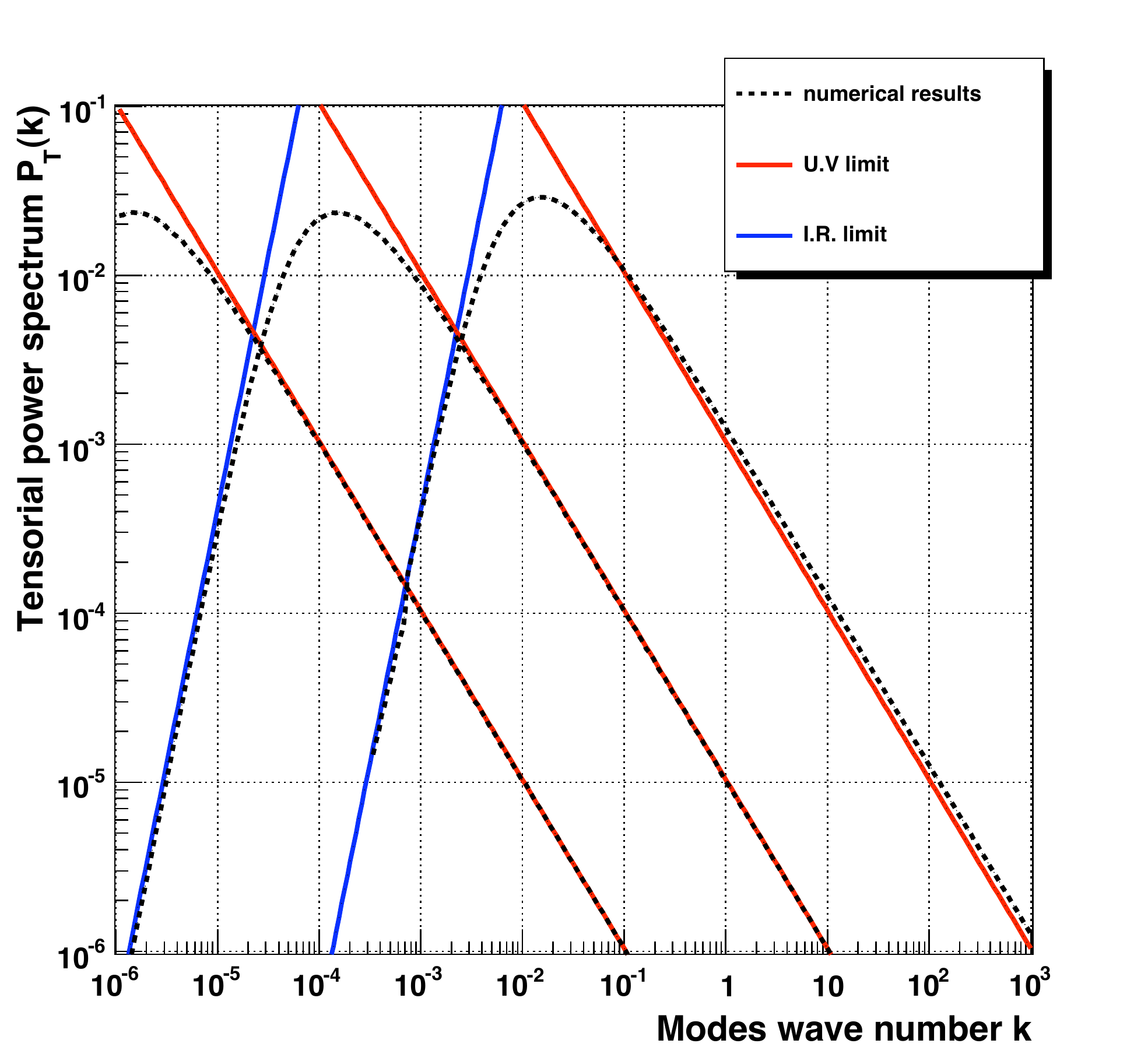}
	\caption{{\it Left panel:} Standard, general relativistic potential (solid line) and holonomy corrected potential (dashed line). {\it Right panel:} Primordial tensor power spectrum, in Planck units, for power-law inflation with $\beta=-2.5$ and $L_{Pl}/L_0=10^{-4}$, $10^{-3}$ and $10^{-2}$  (from left to right): dashed lines are for the numerical results, blue lines are for the IR limit and red lines for the UV limit of the power spectrum. The UV approximation coincides with the primordial spectrum obtained without L.Q.G. correction.}
	\label{figholo}
\end{figure}

A quasi-classical Hamiltonian including holonomy corrections can be derived from the classical and perturbed Hamiltonian by replacing $\bar{k}$ by $\left(\frac{\sin\bar{\mu}\gamma\bar{k}}{\bar{\mu}\gamma}\right)$ \cite{bojo1}. The same replacement has also to be done in the expression of the perturbed extrinsec curvature. This allows to derive an effective equation of motion for gravitational waves containing such a type of quantum corrections. By switching to spatial Fourier space and multiplying $h^a_i$ by the scale factor, one arrives at \cite{grainL.Q.G.2}:
\begin{equation}
	\ddot{\phi}_k+\left(k^2-\frac{\ddot{a}}{a}-V_{holo}(a,\gamma)\right)\phi_k=0,
\end{equation}
where the explicit expression for the potential-like term due to holonomy corrections can be found in \cite{grainL.Q.G.2}. In the case of slow-roll inflation,we have:
\begin{eqnarray}
	a(\eta)&=&L_0\left|\eta\right|^{-1-\epsilon}, \\
	\frac{\ddot{a}}{a}&=&\frac{2+3\epsilon}{\eta^2}, \\
	V_{holo}(a,\gamma)&\propto&-\frac{L_{Pl}}{L_0}\left|\eta\right|^{2\epsilon-2}.
\end{eqnarray}
The $L_0$ quantity is roughly given by the inverse of the energy scale of inflation. Standard quantum mechanical reasoning can be used by interpreting $k^2$ as the energy term, $\ddot{a}/a$ as the potential due to the expanding spacetime and $V_{holo}$ as the potential due to holonomy corrections. Amplification of the gravitational waves will start when $k^2<\ddot{a}/a+V_{holo}$. The standard, general relativistic potential ({\it i.e.} $V_{holo}=0$) and the loop quantum corrected one are compared on the left panel of Fig. \ref{figholo}. It shows that with holonomy corrections taken into account, the phase of amplification starts latter than in the standard general relativistic case, meaning that the quantum corrected power spectrum should be suppressed as compared to the standard one.

Except for the $\epsilon=0$ case, the above defined Schr\"odinger equation cannot be solved analytically. However, the primordial power spectrum can be derived in the infrared ($k\to 0$) and ultraviolet regimes ($k\to\infty$):
\begin{equation}
	P_T^{(IR)}(k)=C\left(\frac{L_{Pl}}{L_0}\right)^{-3/\epsilon}k^3
\end{equation}
and 
\begin{eqnarray}	
	P_T^{(UV)}(k)&=&\left(\frac{L_{Pl}}{L_0}\right)^2\left(\frac{2^{3/2}\Gamma(3/2+\epsilon)}{\pi}\right)^2k^{-2\epsilon} \\
	&=&P_T^{standard}(k)\nonumber
\end{eqnarray}
where $C$ is a numerical constant depending on $\epsilon$ and $P_T^{standard}$ is the standard tensor power spectrum ({\it i.e.} without any Loop Quantum Cosmology corrections). This shows that in the ultraviolet regime, holonomy corrections are subdominant and the tensor power spectrum is equal to the standard, general relativistic prediction in slow-roll inflation. However, holonomy corrections are not negligeable in the infrared part of the spectrum as for very large scales, the power spectrum exhibits now a very blue tilt $n_T=3$, thus confirming our naive reasoning that holonomy corrections should lead to a suppression of the produced gravitational waves. 

Finally, the full power spectrum on the entire $k$-range has been derived using numerical integration. As shown on the right panel of Fig. \ref{figholo}, this confirms the existence of two regimes: at large scales, the power spectrum is very blue because of holonomy corrections whereas at small scales, those corrections are subdominant and the tensor power spectrum is given by the standard, general relativistic red-tilted behaviour. We stress out that those results have been obtained in the general case of power-law inflation, $a\propto\left|\eta\right|^{\beta+1}$ with $\beta\leq-2$. However, this also holds for slow-roll inflation by replacing $(\beta+1)$ by $(-1-\epsilon)$ and then perfoming a Taylor expansion in $\epsilon$.
	
\subsection{Inverse-volume corrections}
\begin{figure}[ht!]
	\includegraphics[scale=0.4]{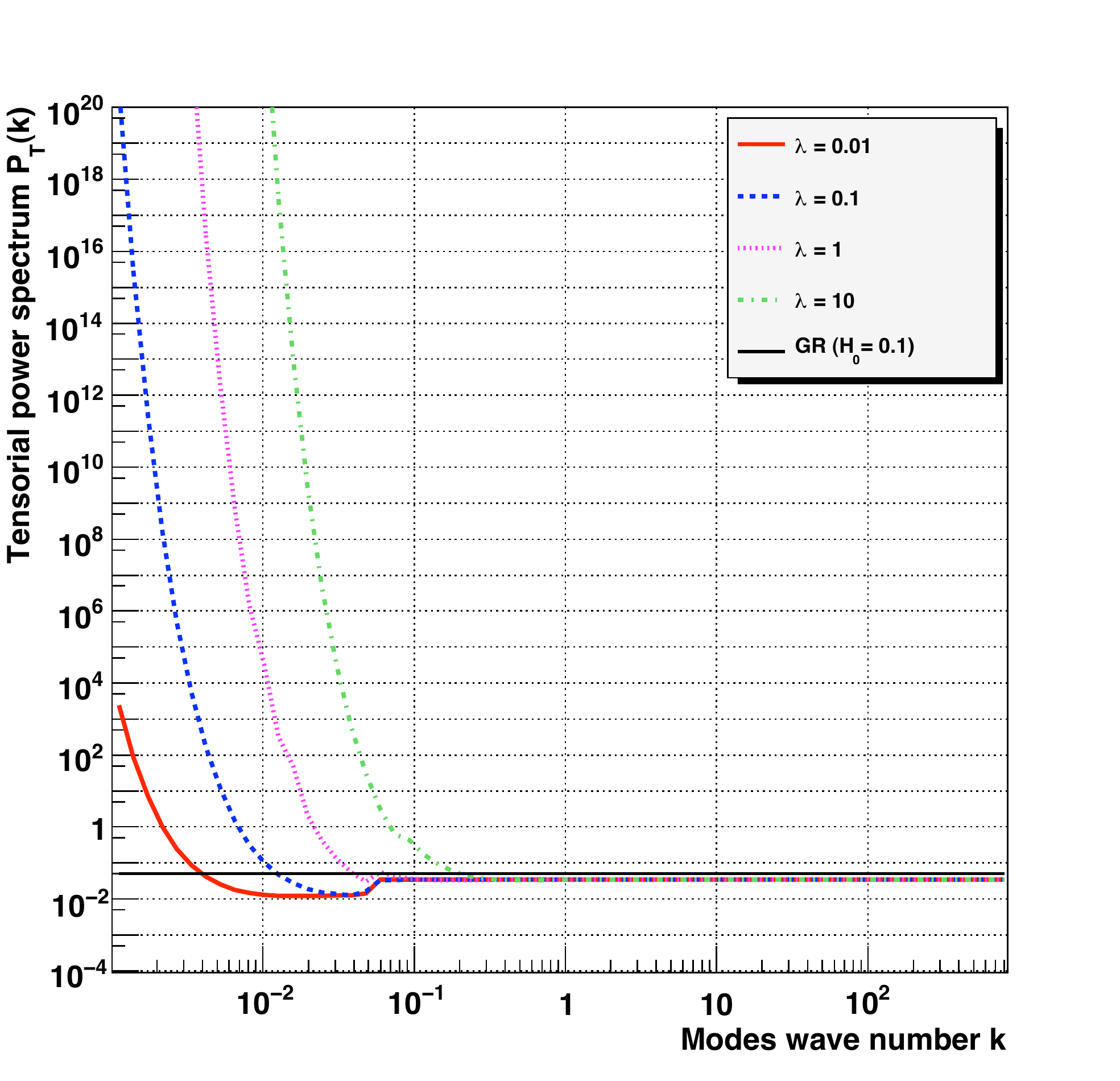}\includegraphics[scale=0.4]{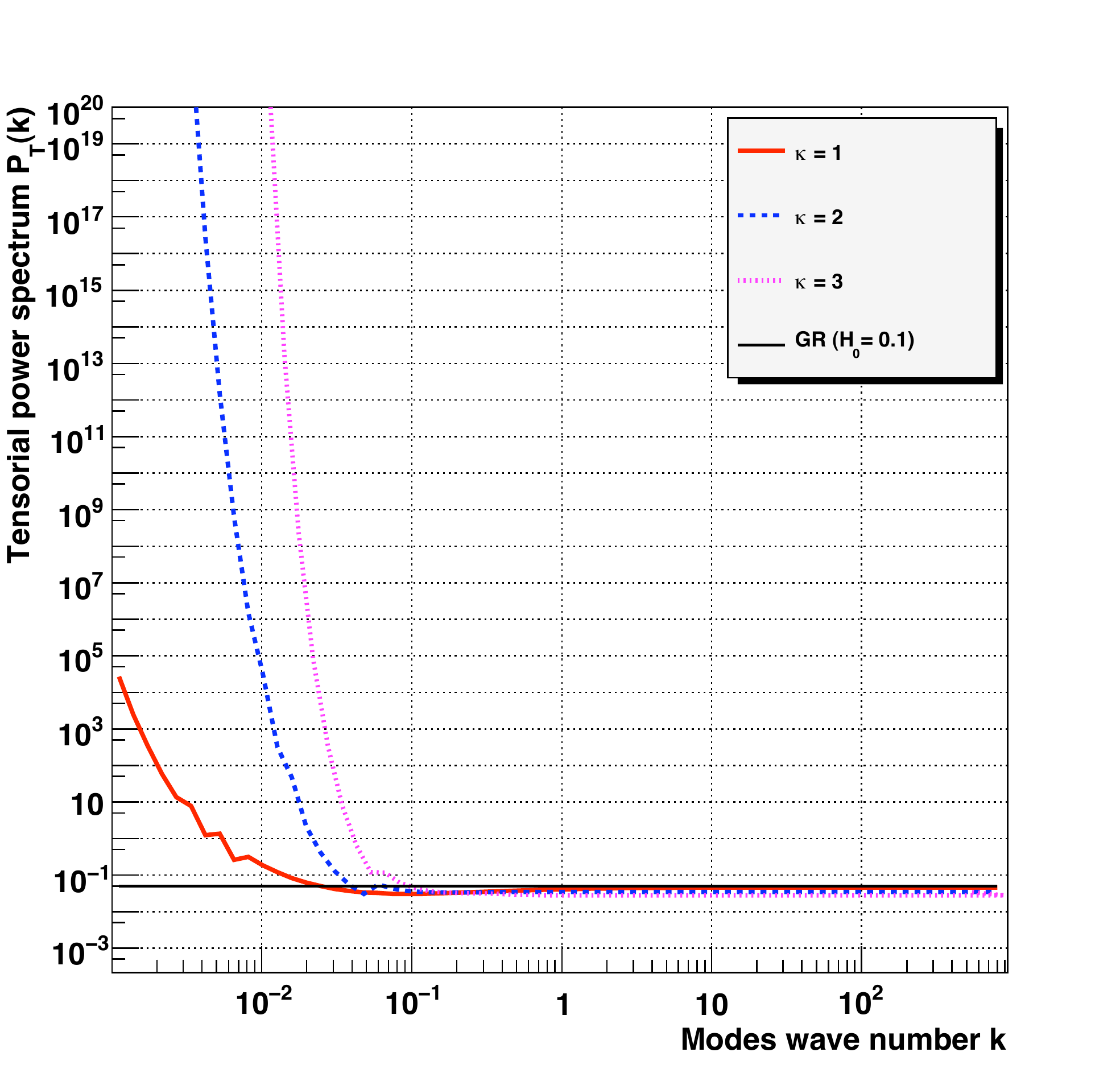}
	\caption{ Primordial tensor power spectrum as a function of the wavenumber for di?erent values of  the L.Q.C. parameters $\lambda$ (left panel) and $\kappa$ (rightpanel).}
	\label{figiv}
\end{figure}

The same approach can also be adopted to include inverse-volume correction \cite{bojo1}. The effective and perturbed gravitational Hamiltonian is now obtained by multiplying the volume element in the classical and perturbed Hamiltonian by the $S$ function. The resulting, Schr\"odinger equation for gravitational waves now reads:
\begin{equation}
	\ddot{\psi}_k+\left(S^2(a)k^2-\frac{\ddot{a}}{a}+\frac{\dot{a}}{a}\frac{\dot{S}}{S}-\frac{3}{4}\left(\frac{\dot{S}}{S}\right)^2+\frac{1}{2}\frac{\ddot{S}}{S}\right)\psi_k=0,
\end{equation}
with
$$
\psi_k(\eta)=\frac{a(\eta)}{\sqrt{S(a)}}\displaystyle\int d^3x {h}^a_i(\eta,\vec{x})e^{i\vec{k}\vec{x}}
$$
and 
$$
S(a)=1+\lambda\left(\frac{L_{Pl}}{a(\eta)}\right)^{\kappa}.
$$
In the slow-roll inflation, the above equation can be analytically solved for $\kappa(1+\epsilon)=2$ by use of Kummer functions (see Sec. II.B. of \cite{grainL.Q.G.3}). The final result is rather involved and in particular makes use of the square modulus of complex valued Euler $\Gamma$ function (see Eq. (73) of \cite{grainL.Q.G.3}). However, the resulting infrared and ultraviolet asymptotic limits are easily expressed using standard transcendal functions:
\begin{eqnarray}
	P_T^{(IR)}(k)&=&\left(\frac{L_{Pl}}{L_0}\right)^2\left(\frac{2^{3/2}\Gamma(3/2+\epsilon)}{\pi}\right)^2\left[2Z(1+\epsilon)\right]^{-3/2-\epsilon}k^3\exp\left[\frac{\pi\sqrt{2Z}(1+\epsilon)}{2k}\right] \\
		&=&P_T^{standard}(k)\times\left[2Z(1+\epsilon)\right]^{-3/2-\epsilon}k^{3+2\epsilon}\exp\left[\frac{\pi\sqrt{2Z}(1+\epsilon)}{2k}\right], \nonumber
\end{eqnarray}
and
\begin{eqnarray}
	P_T^{(UV)}(k)&=&\left(\frac{L_{Pl}}{L_0}\right)^2\left(\frac{2^{3/2}\Gamma(3/2+\epsilon)}{\pi}\right)^2k^{-2\epsilon}\left[1+(3+5\epsilon)\frac{Z}{k^2}+\mathcal{O}(k^{-4})\right] \\
	&=&P_T^{standard}(k)\times\left[1+(3+5\epsilon)\frac{Z}{k^2}+\mathcal{O}(k^{-4})\right], \nonumber 
\end{eqnarray}
with $Z=\lambda(L_{Pl}/L_0)^\kappa$. At small length scales, the tensor power spectrum is poorly affected by inverse-volume corrections which only leads to a very small additional tilt and running, both of them proportionnal to $k^{-2}$. However, at large scales, the production of primordial gravity waves is strongly boosted by the exponential term. We have check thanks to numerical investigations that this strong exponential boost in the infrared regime is a typical feature of the inverse-volume corrections. As shown on Fig. \ref{figiv}, this boost appears for different values of $(\kappa,\lambda)$ and its amplitude scales with the $Z$ parameter.

\subsection{Completion of both corrections}
The two type of corrections implemented in the two previous sections can be treated in a unified way \cite{grainL.Q.G.4}. The complete, quasi-classical and perturbed Hamiltonian of the gravtitational field is derived from the classical and perturbed one by adding the $S$ function and by replacing $\bar{k}$ by $\left(\frac{\sin\bar{\mu}\gamma\bar{k}}{\bar{\mu}\gamma}\right)$ together. The full effective and perturbed Hamiltonian and the extrinsec curvature now reads \cite{grainL.Q.G.4}:
$$
\begin{array}{lll}
	\displaystyle H^{(qc)}_{grav}&\displaystyle=&\displaystyle\frac{1}{2 \kappa}\int_{\Sigma}\mathrm{d}^3x \bar{N} S(\bar{p},\delta E^a_i)\left[-6\sqrt{\bar{p}}\left(\frac{\sin\bar{\mu}\gamma\bar{k}}{\bar{\mu}\gamma}\right)^2 - \frac{1}{2\bar{p}^{3/2}} \left(\frac{\sin\bar{\mu}\gamma\bar{k}}{\bar{\mu}\gamma}\right)^2 (\delta E^c_j\delta E^d_k\delta_c^k\delta_d^j) \right. \nonumber\\
		&&\displaystyle+ \left.\sqrt{\bar{p}} (\delta K_c^j\delta K_d^k\delta^c_k\delta^d_j) - \frac{2}{\sqrt{\bar{p}}} \left(\frac{\sin 2\bar{\mu}\gamma\bar{k}}{2\bar{\mu}\gamma}\right)(\delta E^c_j\delta K_c^j) - \frac{1}{\bar{p}^{3/2}} (\delta_{cd}\delta^{jk}  E^c_j \delta^{ef} \partial_e \partial_f  E^d_k ) \right], \\
	\displaystyle H^{(qc)}_{matter}&\displaystyle=&\displaystyle\displaystyle\int_\Sigma d^3x\left(\frac{1}{2}D(\bar{p})\frac{\dot{\Phi}^2}{\bar{p}^{3/2}}+\bar{p}^{3/2}V(\Phi)\right) \\
	\displaystyle \delta K_a^i &\displaystyle=& \displaystyle\frac{1}{2 S} \dot{h}_a^i + \frac{1}{2} \left( \frac{sin(2 \bar\mu \gamma \bar{k})}{2 \bar\mu \gamma} \right)h_a^i.
\end{array}
$$
As shown in \cite{grainL.Q.G.4}, this leads to the following Schr\"odinger equation for the primordial gravity waves:
\begin{equation}
	\ddot{\psi}_k+\left\{S^2 k^2- \left(\frac{\ddot{a}}{a}+M^2(a) -\frac{\dot{a}}{a} \frac{\dot{S}}{S}+ \frac{3}{4}\left(\frac{\dot{S}}{S}\right)^2  -\frac{1}{2}\frac{\ddot{S}}{S}\right) \right\}\psi_k=0,
\end{equation}
with
$$
M^2(a)=\frac{1}{8\pi G} \frac{\rho_\Phi}{\rho_c}  a^2 \left( \frac{2}{3}\rho_\Phi -\frac{\dot{\Phi}^2}{D(q) a^2} \left( 1- \frac{1}{6} \frac{\dot{D}}{D} \frac{a}{\dot{a}}\right)\right),
$$
and $\psi_k=\frac{a(\eta)}{\sqrt{S(a)}}\int d^3x {h}^a_i(\eta,\vec{x})e^{i\vec{k}\vec{x}}$. Holonomy corrections are encoded in the $M^2$ term whereas inverse-volume corrections are encoded in all the term involving $S$. One can notice that inverse-volume corrections are also involved in the $M^2$ term via the $D$ function because the matter Hamiltonian is also affected by inverse-volume corrections. 

If both corrections are considered, the equation of motion is analytically solved for $\kappa=2$ and $\epsilon=0$ which leads to the following infrared and ultraviolet behaviour \cite{grainL.Q.G.4}:
\begin{eqnarray}
	P_T^{IR}(k) &=& 16 \pi^3 \left(\frac{L_{PL}}{L_0} \right)^2 (Z(1-4\gamma^2\frac{L^2_{Pl}}{L^2_0}))^{-\frac{3}{2}} k^3 \exp\left[{\pi \sqrt{\frac{Z}{8}}\frac{(1-4\gamma^2\frac{L^2_{Pl}}{L^2_0})}{k}}\right] \\
	&=&P^{standard}_T(k)\times(Z(1-4\gamma^2\frac{L^2_{Pl}}{L^2_0}))^{-\frac{3}{2}} k^3 \exp\left[{\pi \sqrt{\frac{Z}{8}}\frac{(1-4\gamma^2\frac{L^2_{Pl}}{L^2_0})}{k}}\right] , \nonumber \\
	P_T^{UV} (k) &= &16 \pi^3 \left(\frac{L_{PL}}{L_0^2} \right)^2 \left[1 + \frac{3}{2} \frac{Z}{k^2} (1 - 4 \epsilon)+\mathcal{O}(k^{-4})\right] k^{-4 \gamma^2L^2_{Pl}/3L^2_0} \\
	&=&P^{standard}_T(k)\times\left[1 + \frac{3}{2} \frac{Z}{k^2} (1 - 4 \epsilon)+\mathcal{O}(k^{-4})\right] k^{-4 \gamma^2L^2_{Pl}/3L^2_0}. \nonumber
\end{eqnarray}
This specific case shows that at large length scales, inverse-volume corrections is the dominant contribution leading to the strong exponential boost. However, at small scales, though both corrections are subdominant as compared to the standard result, holonomy and inverse-volume corrections have the same {\it amplitude}, each of them leading to a small additional tilt since the $k^{-4 \gamma^2L^2_{Pl}/3L^2_0}$ is due to holonomy corrections and the $\left[1 + \frac{3}{2} \frac{Z}{k^2} (1 - 4 \epsilon)+\mathcal{O}(k^{-4})\right]$ is due to inverse-volume corrections.

\section{Conclusion}
Testing quantum theories of gravity is probably one of the most important challenge of current fundamental physics. Loop Quantum Gravity has not yet been experimentally probed but it seems that cosmological observations could allow for a clear signature of L.Q.G. effects. In this communication, we have shown how the tensor power spectrum is affected by L.Q.G. corrections via a modified dispersion relation for the propagation of primordial gravity waves in a inflationary background. This constitutes a first step towards a full derivation of the statistical properties of the cosmological perturbations produced in the early universe where L.Q.G. corrections on the background dynamics are also included. Although B-mode detection could be achieved by the PLANCK satellite and is the main goal of several dedicated experiments for the forthcoming decade, the signal remains very difficult to extract from the lensing background and some further refinements are required to quantify the amplitude of the expected L.Q.G. effects. Finally, the formalism established in this article should be used for L.Q.G. corrections to scalar perturbations that are just being investigated \cite{bojo3} and could be even more promising from the observational viewpoint.

\section{Acknowledgments}
The author would like to thank his collaborators  A. Barrau, T. Cailleteau and A. Gorecki, in Grenoble, as well as J. Mielczarek in Cracaw. He is also grateful to M. Bojowald for stimulating discussions on Loop Quantum Cosmology during his stay in Paris. This work was partially supported by Hublot-Gen\`eve company and by the European Commission Marie Curie IR Grant, MIRG-CT-2006-036614

\end{document}